\documentclass[12pt]{iopart}

\usepackage{graphicx}

\def\alt{\mathrel {\mathpalette \vereq <}}
\def\agt{\mathrel {\mathpalette \vereq >}}
\def\mathpalette#1#2{\mathchoice {#1\displaystyle {#2}}%
                                 {#1\textstyle {#2}}%
                                 {#1\scriptstyle {#2}}%
                                 {#1\scriptscriptstyle
{#2}}}
\def\vereq#1#2{\lower 3pt\vbox {\baselineskip 1.5pt \lineskip 1.5pt \ialign
{$\m@th #1\hfill ##\hfil $\crcr #2\crcr \sim \crcr }}}
\def\m@th{\mathsurround=0pt}

\begin{document}
\title{Spin-fluctuation drag thermopower
 of nearly ferromagnetic metals
}

\author{Takuya Okabe}
\address{
Faculty of Engineering, Shizuoka University, 
3-5-1 Johoku, 
Hamamatsu 432-8561,Japan}


\begin{abstract}
We investigate theoretically the Seebeck effect 
in materials close to a ferromagnetic quantum critical point
to explain anomalous behaviour 
at low temperatures.
It is found that the main effect of spin fluctuations
is to enhance 
the coefficient of the leading $T$-linear term,
and a quantum critical behaviour 
characterized by a spin-fluctuation temperature 
appears in the temperature 
dependence of correction terms
as in the specific heat.
\end{abstract}





%


\maketitle

\section{Introduction}

Experiments on clean materials 
near ferromagnetic quantum critical point
(QCP)
have revealed unusual properties, including 
non Fermi liquid transport
and unconventional superconductivity.\cite{saxena00,nblkstbbf05}
The effects caused by
quantum critical dynamics of spin fluctuations
on the specific heat coefficient, 
the spin susceptibility, the resistivity, and so on,
have been elucidated analytically at low temperatures.\cite{de66,bmf68,mathon68,moriya85,lrvw07}
In most of such theoretical analyses made so far,
critical spin fluctuations are regarded to stay in thermal equilibrium.
On the other hand, one may conceive of 
its inequilibrium counterpart of anomalous behaviours as well,
which would be of fundamental interest too and 
should be paid due attention theoretically.
As a representative of such phenomena, 
there are 
observations
suggesting spin-fluctuation (or paramagnon) drag thermopower.
In the Seebeck coefficient $S(T)$ of ${\rm UAl_2}$, for example, 
there have remained a structure at low temperature,
which is observed experimentally,\cite{afss79,po97}
but left unexplained
theoretically.\cite{ijc78,cij78} 
Among others, the most typical
clear-cut experimental evidence would be those reported 
by Gratz $et$ $al$.,\cite{grbbmg95,gratz97,gm01}
where the pronounced low-temperature minimum in $S(T)$
of strong paramagnet
$\rm RCo_2$ (R$=$Sc, Y and Lu) was attributed to the paramagnon
drag effect.
Recently, Matsuoka $et$ $al$.\cite{mhittm05} found 
a similar structure for $\rm A Fe_4Sb_{12}$ (A$=$Ca, Sr and Ba).
In effect, 
Takabatake $et$ $al$.\cite{tmnhmsus06} 
made it clear that
the anomaly in $S(T)$ 
is indeed caused by the ferromagnetic spin fluctuations
prevalent in the materials
by showing that those structure is completely suppressed by applying a
uniform magnetic field.
In contrast with the accumulating experimental evidence,
%
there seems no 
theory 
to compare with the experiments available so far,
with the exception of a brief account 
on 
a qualitative effect expected for {\it localized} spin
fluctuations around impurity sites of alloys.\cite{kaiser76}
In this paper, 
we discuss an effect of uniform spin fluctuations in a
translationally invariant system,
and intend to provide a more solid footing on which to discuss the phenomenon.

In section~\ref{sec:model}, 
we give an outline of a two-band model, which we adopt as a relevant model,
along with 
approximations and assumptions conventionally made.
In section~\ref{sec:sfd}, 
we introduce a function $\Phi_k^d$ to represent 
inequilibrium displacement of 
spin fluctuations.
In section~\ref{sec:ltc}, we discuss that
the leading effect of spin fluctuations appears on the $T$-linear term
of $S(T)$.
In effect, in section~\ref{subsec:ee},
we discuss that the leading term contribution follows
a universal relation to the specific heat, that is, $ q\equiv e S/ C \simeq \pm 1$
revealed by Behnia $et$ $al$.\cite{BJF04}
In the higher order terms,
we have to consider 
not only a critical effect originating from equilibrium quantities,
but also a genuinely non-equilibrium effect which has not been investigated before.
In section~\ref{sec:std}, 
we investigate the latter contributions to find
a characteristic temperature dependence,
and the results are 
summarized in the last subsection~\ref{sec:summary}. 
In section~\ref{sec:disscs}, we discuss the results
and comparison is made with experiment.

\section{\label{sec:model}Model}

Let us introduce a two-band paramagnon model, which is conventionally employed
to explain an enhanced resistivity of transition metals at low temperature.\cite{mathon68,ml66,rice68}
%
The model has been applied successfully to explain, e.g., 
a saturation behaviour at elevated temperatures
by taking into account a proper temperature dependence of spin susceptibility.\cite{jbc74,um75}

The model is comprised of two types of electrons, 
i.e., wide-band conduction electrons
and narrow-band itinerant electrons 
on the border of ferromagnetism.
We denote the former as the $s$ electron and the latter as the $d$
electrons, representatively.
The Hamiltonian consists of three parts, 
\[
 H=H_s + H_{sd} + H_d.
\]
The free Hamiltonian of the $s$ electron is given by 
\[
 H_s =\sum_{k\sigma} \varepsilon_s (k)  c^\dagger_{k\sigma} c_{k\sigma},
\]
where 
$c^\dagger_{k\sigma}$ and $c_{k\sigma}$
are the creation and annihilation operators for 
the electron with momentum $k$ and spin $\sigma$.
For simplicity,
it is often assumed 
that
the $s$ electrons make a parabolic band with mass $m_s$, i.e., 
\begin{equation}
 \varepsilon_s (k) =\frac{k^2}{2m_s}.
\label{varepsk} 
\end{equation}
At each site $i$, 
they 
are scattered by 
the spin ${\bi S}_i$ of the $d$ electron at the same site
through the Kondo $s$-$d$ coupling,
\begin{equation}
 H_{sd} = J \sum_i {\bi s}_i\cdot {\bi S}_i,
\label{Hsd}
\end{equation}
where $J$ denotes a coupling constant,
and 
${\bi s}_i = \frac{1}{2}\sum_{\sigma \sigma'} c^{\dagger}_{i\sigma} 
{\bi \tau}_{\sigma\sigma'}c_{i\sigma}$
is the spin 
of the $s$ electron at the site $i$ expressed in terms of 
the Pauli matrix vector ${\bi \tau}_{\sigma\sigma'}$.
Similarly, 
the $d$ electron spin at the site $i$ is given by
${\bi S}_i = \frac{1}{2}\sum_{\sigma \sigma'} d^{\dagger}_{i\sigma} 
{\bi \tau}_{\sigma\sigma'}d_{i\sigma}$
in terms of the creation and annihilation operators $d^\dagger_{i\sigma}$ and $d_{i\sigma}$ 
for the $d$ electron.
Spin dynamics of the $d$ electrons
is described by the Hubbard Hamiltonian,
\begin{equation}
 H_d =\sum_{k\sigma} \varepsilon_d (k)  d^\dagger_{k\sigma} d_{k\sigma} 
+ U \sum_i n_{i\uparrow}n_{i\downarrow},
\label{Hd}
\end{equation}
where $n_{i\sigma}=d^\dagger_{i\sigma} d_{i\sigma}$ $(\sigma
=\uparrow,\downarrow)$ is the number operator
of the $d$ electron at the site $i$.
The on-site repulsion $U$ is fixed such that
the $d$ band is nearly ferromagnetic.
To make analytical evaluation feasible,
it is often assumed further that
the $d$ electrons are also parabolic with a different mass $m_d$
heavier than $m_s$, i.e., 
\begin{equation}
 \varepsilon_d (k) =\frac{k^2}{2m_d},
\label{varepdk} 
\end{equation}
and $m_d \gg m_s$.
The latter inequality is regarded
as the basic ingredient of the model.
Hence the $d$ electrons act as heavy and fluctuating scatterers 
against the $s$ electrons
through the coupling of (\ref{Hsd}).
In effect, this is taken into account as 
the second order effect 
with respect to the coupling $J$, 
i.e., through the Born approximation.\cite{ml66}
Then, the $d$ electron 
comes into play through the 
(transverse) spin susceptibility
$\chi(q, \omega)$. 
%
In the random phase approximation, it is given by
\begin{equation}
 \chi(q,\omega) 
=\frac{ \chi_{0}(q,\omega) }{
1-U  \chi_{0}(q,\omega)},
\label{chi+-qom}
\end{equation}
where
\begin{equation}
\chi_{0}(q,\omega) = \sum_k \frac{f^0_k -f^0_{k+q}}{\varepsilon_d(k+q)
-\varepsilon_d(k) - \omega -{\rm i}\delta
}.
\label{chi0qomega}
\end{equation}
Here, $f_k^0 \equiv f^0(\varepsilon_d (k))
=1/({\rm e}^{(\varepsilon_d (k)-\mu)/T}+1)
$
is the Fermi distribution function,
and $\delta$ is a positive infinitesimal.
To investigate 
critical properties at low temperatures,\cite{ikk63}
(\ref{chi0qomega}) is expanded 
for small $q$ and $\omega/q$
as
\begin{equation}
\chi_{0}(q,\omega) =  
\rho_{F,d}
\left( 1-\frac{1}{12} \bar{q}^2 +{\rm i} \frac{\pi}{4} \frac{\bar{\omega}}{\bar{q}}
\right),
\label{chi0qom}
\end{equation}
for $\bar{\omega}< 2 \bar{q} $, 
where $\bar{q} = q/k_{F,d}$
and  $\bar{\omega} = \omega/\varepsilon_{F,d}$
are the momentum and energy normalized by the Fermi momentum $k_{F,d}$
and the Fermi energy $\varepsilon_{F,d}$ of the $d$ electron.
$\rho_{F,d}
=m_d k_{F,d}/2\pi^2$
is the density of states (DOS) at the Fermi level 
of the $d$ electron per spin.
Substituting
(\ref{chi0qom}) into 
(\ref{chi+-qom}),
we obtain
\begin{equation}
 \chi
(q,\omega)  = \frac{\rho_{F,d}}{K_0^2 + \frac{\bar{U}}{12}\bar{q}^2
  -{\rm i} \frac{\pi\bar{U}}{4} \frac{\bar{\omega}}{\bar{q}}}, 
\label{chi+-qomega}
\end{equation}
for $\bar{\omega}< 2 \bar{q} $,
where $\bar{U}=\rho_{F,d}U$, and $K_0^2= 1-\bar{U} (\ll 1)$
represents the distance to the 
QCP.
%

The intrinsic transition probability $ {\cal Q}_{k,q}^{k+q}$
that an $s$ electron with momentum $k$ 
is scattered to $k+q$
by absorbing a spin fluctuation with $q$ and $\omega$
via the coupling in (\ref{Hsd})
is given by 
\begin{equation}
{\cal Q}_{k,q}^{k+q}(\omega) = 
\frac{3J^2}{4} S(q,\omega),
\end{equation}
where 
$S(q,\omega)$ denotes the Fourier transform of 
the spin density correlation function, 
which is related to the dynamical susceptibility 
by the fluctuation dissipation theorem.\cite{ikk63}
\begin{equation}
S(q,\omega)
=\frac{2 }{1-{\rm e}^{-\omega/T}}{\rm Im}\chi
(q,\omega).
\end{equation}
The equilibrium transition rate is given by 
\begin{equation}
\fl
{\cal P}_{k,q}^{k+q}=
\int {\rm d}\omega 
(1-f^0(\varepsilon_s({k+q}))) f^0(\varepsilon_s(k)) n^0(\omega)
 {\cal Q}_{k,q}^{k+q}(\omega)
\delta(\omega +\varepsilon_s(k)-\varepsilon_s({k+q})), 
\label{calP=intdomega1-f0fn0Q}
\end{equation}
where
$f^0(\varepsilon_s(k))$ is the Fermi factor for the $s$ electron, and
$n^0(\omega) =1/({\rm e}^{\omega/T}-1)$ is the Bose function.
With this ${\cal P}_{k,q}^{k+q}$,
transport coefficients are derived 
by following the formal transport theory of Ziman\cite{ziman60}
(cf. \ref{sec:formaltransporttheory}).
Transport properties of the $s$ electrons
in an electric field ${\bi E}$ and a gradient of temperature $\nabla T$
are described by the Boltzmann transport equation,
\begin{equation}
 -{\bi v}_s(k) \cdot \nabla T  \frac{\partial f^0(\varepsilon_s(k))}{\partial T}
 -e {\bi v}_s(k) \cdot {\bi E} \frac{\partial
f^0(\varepsilon_s(k))}{\partial \varepsilon_s(k)}
= 
- \dot{f}_k,
\label{boltzeq0}
\end{equation}
where
\(
{\bi v}_s(k) =\nabla_k \varepsilon_s(k)
\) is the velocity of the $s$ electron,
and $e (<0)$ is the electronic charge.
The right-hand side in (\ref{boltzeq0})
is the collision integral for the $s$ electron.

To linearize the transport equation
for the conduction electrons,
a function $\Phi_k^s $ to represent
the displacement of the distribution function $f(\varepsilon_s(k))$
from the equilibrium one $f^0(\varepsilon_s(k))$
is introduced, i.e.,  by
\begin{equation}
 f(\varepsilon_s(k)) =f^0(\varepsilon_s(k))  
-\frac{\partial f^0(\varepsilon_s(k)) 
}{\partial \varepsilon_s(k)}
\Phi_k^s.
\label{f=f0-Phik} 
\end{equation}
On the contrary,
the $d$ electrons are 
commonly assumed to 
stay in equilibrium, despite the applied fields.
Then, for the collision integral in (\ref{boltzeq0}),
we obtain
\begin{equation}
\dot{f}_k  =
-\frac{1}{T}\sum_{q}  
\left(
\Phi_k^s - \Phi_{k+q}^s\right)
{\cal P}_{k,q}^{k+q}.
\label{fkscattLinear}
\end{equation}
%

For definiteness, 
let the fields ${\bi E}$ and $\nabla T$ be in the direction parallel to 
a unit vector ${\bi u}$.
For the isotropic model,
the magnitudes of the electric and heat currents due to the $s$ electrons 
are given by 
\begin{equation}
 {J}_s[\Phi^s] = 2e \sum_k 
{\bi u}\cdot {\bi v}_s(k)
\left(
-\frac{\partial f^0(\varepsilon_s(k)) }{\partial \varepsilon_s(k)}
\right)
 \Phi_k^s,
\label{J=esumkvck}
\end{equation}
and 
\begin{equation}
 {U}_s[\Phi^s] =  2\sum_k {\bi u}\cdot {\bi v}_s(k)
(\varepsilon_s(k) -\mu)
\left(
-\frac{\partial f^0(\varepsilon_s(k)) 
}{\partial \varepsilon_s(k)}
\right)
 \Phi_k^s.
\label{U0=sumkvck}
\end{equation}
The factor 2 in front of the $k$ sum accounts for the two spin
components.
As noted below (\ref{f=f0-Phik}),
it is conventionally assumed that
the corresponding currents due to the $d$ electrons
are neglected against the $s$ electron currents.

To obtain a solution $\Phi_k^s$,
one may set $\Phi_k^s =\tau {\bi u}\cdot {\bi v}_s(k)$,
while the constant $\tau$ is fixed by the equation.
Consequently,
for the electric resistivity $R=R_0(T)$
and the diffusion thermopower coefficient $S=S_0^s(T)$, 
we obtain 
\begin{equation}
 R_0 (T)= \frac{P_{ss}}{({J_s}[\Phi^s])^2},
\label{R0T=fracpss}
\end{equation}
and
\begin{equation}
 S_0^s(T) =
\frac{1}{T}
\frac{{ U_s}[\Phi^s]}{{J_s}[\Phi^s]},
\label{S0sT}
\end{equation}
where
\begin{equation}
P_{ss}  =\frac{1}{T}\sum_{k,q}
{\cal P}^{k+q}_{kq}  \left(
 \Phi_k^s 
- 
\Phi_{k+q}^s
\right)^2.
\end{equation}
%
%
%
%
%
%
The ordinary diffusion thermopower in (\ref{S0sT}) 
is linear in $T$ at low temperature, and
is often expressed as 
\[
S_0^s (T)
=\frac{\pi^2 T}{3e}  \frac{\partial \log
\sigma_s(\varepsilon_{F,s}) }{\partial \varepsilon},
\]
in terms of the spectral conductivity $\sigma_s(\varepsilon)$
of the conduction electron.

\section{\label{sec:sfd}
Spin-fluctuation drag}

As remarked above,
the $d$ electrons are customarily assumed to stay in equilibrium 
regardless of the applied fields.
%
To generalize the above framework to describe 
spin fluctuations with a shifted distribution theoretically,
let us consider a bare dragged susceptibility $\chi^{q_0}_0(q,\omega)$, which is obtained 
by shifting uniformly the equilibrium bare susceptibility $\chi_0(q,\omega)$
in (\ref{chi0qomega})
by a small 
but finite amount
${\bi q}_0$ in momentum space.
Similarly, we may define $\chi^{q_0}(q,\omega)$ for 
the full susceptibility $\chi(q,\omega)$ as well.
Hence, $\chi^{q_0}(q,\omega)$ is strongly peaked at ${\bi q}={\bi q}_0$.

First we derive a simple relation between $\chi^{q_0}_0(q,\omega)$ and $\chi_0(q,\omega)$.
According to (\ref{chi+-qom}),
we will obtain a similar relation for the full susceptibility.
For the derivation, we introduce a shifted energy 
of the $d$ electron,
\begin{equation}
 \varepsilon_d^{q_0}(k)
= \varepsilon_d (k-q_0)
\simeq \varepsilon_d(k) - {\bi q}_0\cdot{\bi v}_{d}(k),
\label{espdq0k} 
\end{equation}
where ${\bi v}_{d}(k) = \nabla_k \varepsilon_d$.
Then, $\chi^{q_0}_0(q,\omega)$ is obtained 
by distributing 
the $d$ electron with momentum $k$
according to 
the shifted distribution $f^0(\varepsilon^{q_0}_{k})$\footnote{
A similar consideration was taken to derive
the Drude weight of a Fermi liquid.\cite{cond1}}, that is to say, 
by 
\begin{eqnarray}
 \chi^{q_0}_0(q,\omega) 
&=& \sum_k \frac{
f^0(\varepsilon_d^{q_0}({k}))-f^0(\varepsilon_d^{q_0}({k+q}))
}{\varepsilon_d(k+q)
-\varepsilon_d({k})-\omega}
\label{chiq00}
\\
&=&\sum_k 
\frac{
f^0(\varepsilon_d({k}))-f^0(\varepsilon_d({k+q}))
}{\varepsilon^{-q_0}_d(k+q)
-\varepsilon^{-q_0}_d(k) - \omega
}.
\end{eqnarray}
Thus, by (\ref{espdq0k}),
we obtain the relation
\begin{equation}
\chi_0^{q_0} (q,\omega) \simeq
\chi_0(q,\omega + 
{\bi q}_0\cdot{\bi v}_d(q)).
\label{chi0q0qom=chi0qom-q0} 
\end{equation}
This is the result on which we base ourselves in the following.

According to (\ref{chi0q0qom=chi0qom-q0}),
the
drag effect is 
described by a function $\Phi_q^d \equiv{\bi q}_0\cdot{\bi v}_{d}(q)$.
To understand what this represents,
it is instructive to consider the isotropic case of
(\ref{varepdk}), where ${\bi v}_{d}(q)= {\bi q}/m_d$. 
In this case, we obtain $\Phi_q^d={\bi V}\cdot{\bi q}$ 
where ${\bi V}={\bi q}_0/m_{d}$ denotes 
a uniform drift velocity
of the $d$ electrons, or the spin fluctuations.
In effect, the energy $\varepsilon^{-q_0}_d(k)$
represents
the excitation energy of the $d$ electron 
in the moving frame drifting 
with the velocity ${\bi V}$.
This is just a Galilean transformation.
Indeed, noting that we can write
\[
f^0(\varepsilon_d^{q_0}({k})) 
=f^0(\varepsilon_d({k}))
-\frac{\partial f^0(\varepsilon_d(k)) 
}{\partial \varepsilon_d(k)}
\Phi_k^d,
\]
and comparing this with (\ref{f=f0-Phik}),
it would be clear that
the new function $\Phi_k^d$ 
represents the distribution shift
of the $d$ electrons,
just as  $\Phi_k^s$ does for the $s$ electrons.
Thus, we argue that
the drag effect of spin fluctuations
is described in terms of $\Phi_q^d$ in the way that
$\chi_{\rm drag} (q,\omega)$ of the dragged fluctuations 
is represented as
\begin{equation}
\chi_{\rm drag} (q,\omega) =
\chi(q,\omega+\Phi_q^d),
\label{chiphiq=}
\end{equation}
in terms of the equilibrium susceptibility $\chi(q,\omega)$.

Given the above argument, 
we have next to investigate how the formalism 
in the last section should be affected
by a non-vanishing $\Phi_q^d$. 
The first effect is to modify the collision integral in
(\ref{fkscattLinear}).
To see this, here we follow
how (\ref{fkscattLinear}) is derived.
The collision term
in the right-hand side of 
(\ref{boltzeq0}) is explicitly given by 
\begin{eqnarray}
\fl 
\dot{f}_k=
-\sum_{q} 
\int {\rm d}\omega 
\left[
(1-f_{k+q}) f_k 
n^0(\omega)
-
f_{k+q}(1- f_k)
(n^0(\omega) +1)
\right]
\\
 \times {\cal Q}_{k,q}^{k+q}(\omega)
\delta(\omega +\varepsilon_s(k)-\varepsilon_s({k+q})), 
\label{fkscattFull}
\end{eqnarray}
where we denoted $f_k=f(\varepsilon_s(k))$ for the distribution function.
According to the condition of 
detailed balance,
the equilibrium distribution functions $f^0_k$ and $n^0(\omega)$
satisfy the relation
\begin{equation}
 (1-f^0_{k+q}) f^0_k n^0(\omega)
-
f^0_{k+q}(1- f^0_k)(n^0(\omega) +1)=0.
\label{detailedbalance}
\end{equation}
Accordingly, 
by substituting (\ref{f=f0-Phik})
into (\ref{fkscattFull}),
we obtain (\ref{fkscattLinear})
to the linear order in $\Phi_k^s$.
To go further to take into account the inequilibrium shift of the $d$ electrons,
we regard that
${\cal Q}_{k,q}^{k+q}(\omega)$ in (\ref{fkscattFull}), 
or  ${\cal P}_{k,q}^{k+q}$ of (\ref{calP=intdomega1-f0fn0Q}),
depends on 
$\chi_{\rm drag}(q,\omega)$
in place of $\chi(q,\omega)$.
Then 
we can make use of (\ref{chiphiq=}).
The first effect of $\Phi_q^d$
is to 
change the scattering probability ${\cal P}_{k,q}^{k+q}$,
which eventually has no effect owing to 
(\ref{detailedbalance}).
The second is to replace $n^0(\omega)$  
in (\ref{fkscattFull})  by
\begin{equation}
n^0(\omega-\Phi_q^d)
\simeq
n^0(\omega)  
-\frac{\partial n^0}{\partial \omega}
\Phi_q^d.
\label{nomega=n0omega-}
\end{equation}
As a result, we obtain
\begin{equation}
 \dot{f}_k =
-\frac{1}{T}\sum_{q}  
\left(
\Phi_k^s + \Phi_q^d - \Phi_{k+q}^s
\right)
{\cal P}_{k,q}^{k+q}.
\label{fkscattLinearD}
\end{equation}

At this point, 
 (\ref{fkscattLinearD})
clearly indicates
a close analogy to the similar problem of phonon drag.\cite{ziman60}
On the one hand, we can reproduce the previous results 
under the assumption $\Phi_q^d=0$ of no drag.
%
On the other hand,
owing to $\Phi_q^d$ in (\ref{fkscattLinearD}),
we can recover the correct identity 
$\dot{f}_k =0$
when the model is genuinely isotropic
as implied by (\ref{varepsk}) and (\ref{varepdk}).
In fact, in this case, we may set
\begin{equation}
\Phi_k^s =\Phi_k^d= {\bi u}\cdot {\bi k},
\label{phiks=ukphiqd=uq}
\end{equation}
where we put ${\bi V}={\bi u}$ without loss of generality. 
Then the null result 
for (\ref{fkscattLinearD}) obtains from the total momentum conservation.
This means that, if properly treated, 
the model should give no resistivity at all,
irrespective of strong scatterings with spin fluctuations.
In effect, 
the spin fluctuations 
in the inequilibrium state represented by (\ref{phiks=ukphiqd=uq})
are completely dragged
along with the conduction electron currents.
It is the fully dragged state in which
all the $s$ and $d$ electrons drift with the same uniform velocity 
${\bi V}$,
independently of the electric field ${\bi E}$.
This 
is the opposite limit to the case $\Phi_q^d=0$ without drag.
In practice, 
in any case, we should have 
a finite rate 
$\dot{f}_k $
by some mechanism neglected in the simple model, 
e.g., by Umklapp scatterings or by scatterings with extraneous agents.
Moreover, generally, 
in order to investigate the degree of drag quantitatively, e.g., 
the temperature dependence through 
a wide range over a characteristic spin fluctuation temperature,
$\Phi_q^d$ should be determined consistently on the
basis of its own transport equation.
In general, the $k$ dependence of $\Phi_k^s$ and $\Phi_k^d$ may not be as simple as 
in (\ref{phiks=ukphiqd=uq}). 

%
For definiteness, however,
we restrict ourselves to the low temperature regime,
where 
we make use of 
the full drag assumption, (\ref{phiks=ukphiqd=uq}),
to elucidate non-trivial effects arising from 
our extra degree of freedom, $\Phi_q^d$.
%
%
%
A formal theory to discuss a general case is given in \ref{sec:formaltransporttheory}.

\section{\label{sec:ltc}Leading effect}

\subsection{Limiting cases}

%
%

In the original model, 
the $d$ electron currents are neglected 
on the basis of 
the basic inequality $|{\bi v}_s(k)| \gg |{\bi v}_d(k)|$, 
or $m_d \gg m_s$.\cite{ziman60,mott35}
Close inspection indicates that this is concluded
%
through the additional implicit assumption 
$\Phi_k^i ={\bi u}\cdot {\bi v}_i(k)$ $(i=s,d)$
on the solutions of the transport equations,
namely, by $\Phi_k^s\gg \Phi_k^d\simeq 0$.
%
%
As we saw above in (\ref{phiks=ukphiqd=uq}),
%
this does not hold true 
in the presence of 
the $d$ electron drag.
In effect, 
the leading term contribution to the thermopower will arise from 
those dragged $d$ electron currents,
which would outweigh the normal diffusion term $S_0^s(T)$ 
in (\ref{S0sT}) due to the conduction electrons 
by a factor of $m_d/m_s\gg 1$. 

%

We obtain from (\ref{varepsk}), (\ref{J=esumkvck}), 
and $\Phi_k^s ={\bi u}\cdot {\bi k}$, 
\begin{equation}
J_s^s\equiv {J}_s [ \Phi^s] 
= \frac{2 e }{3} v_{F,s}k_{F,s}\rho_{F,s},
\label{JssequivJsphis}
\end{equation}
where $v_{F,s}=k_{F,s}/m_s$ is the Fermi velocity.
Similarly, 
(\ref{U0=sumkvck}) may be written as 
\begin{equation}
U_s^s\equiv {U}_s [ \Phi^s] =
 \frac{\pi^2}{3e} T^2
 \frac{\partial {J}_s^s }{\partial \varepsilon_{F,s}}.
\end{equation}
where 
$\varepsilon_{F,s}$ 
is the Fermi energy.
The latter is obtained by 
expanding 
the integrand in (\ref{U0=sumkvck})
with respect to the excitation energy
$\varepsilon_s(k) -\mu$.
The factor of ${\pi^2}T^2/3$ derives
from the energy integral over $\varepsilon_s(k)$ to replace the $k$ sum.
Hence, from (\ref{S0sT}) we obtain
the ordinary $T$-linear Seebeck coefficient
\begin{equation}
 S_0^s=
 \frac{\pi^2}{3e } 
 \frac{\partial \log {J}_s^s }{\partial \varepsilon_{F,s}} T.
\label{S0^s}
\end{equation}
In the same manner, the $d$ electron currents are evaluated.
We may use
\begin{equation}
J_d^d\equiv  {J}_d[\Phi^d] = 2e \sum_k 
{\bi u}\cdot {\bi v}_d(k)
\left(
-\frac{\partial f^0(\varepsilon_d(k)) }{\partial \varepsilon_d(k)}
\right)
 \Phi_k^d,
\label{J=esumkvdk}
\end{equation}
in place of (\ref{J=esumkvck}),
and $U_d^d\equiv {U}_d[\Phi^d]$ as in (\ref{U0=sumkvck}),
with which we obtain
\begin{equation}
 S_0^d\equiv \frac{ U_d^d}{T{J_d^d}}=
 \frac{\pi^2}{3e } 
 \frac{\partial \log {J}_d^d }{\partial \varepsilon_{F,d}} T,
\label{S0^d}
\end{equation}
as in (\ref{S0^s}).
Formally, this represents
the diffusion thermopower due to the $d$ electrons,
as $S_0^s$ 
does for the $s$ electrons.
Therefore, we should expect
\begin{equation}
|S_0^d| \gg |S_0^s|,
\label{s0dggs0s}
\end{equation}
for $S_0^i$ is proportional to the mass $m_i$.
Still, it is remarked that
$S_0^d$ in (\ref{S0^d}) is not a directly observable quantity in general.
In fact, from (\ref{S=frac1TUs+Ud/Js+Jd}),
the total thermopower is given by 
\begin{equation}
S_0
= \frac{
U_s^s+U_d^d
}{T
\left(
J_s^s
+
J_d^d
\right)
}.
\label{S=fracUss+Udd/T}
\end{equation}
Therefore, 
on the one hand, in the conventional case without $d$ electron drag, where 
$|J_s^s|\gg |J_d^d|$ and $|U_s^s|\gg |U_d^d|$,
we recover the normal result $S_0\simeq S_0^s$.
On the other hand, in the opposite limiting case of the full drag,
as the two currents $J_s^s$ and $J_d^d$ become comparable with each other,
we expect a sizable modification from the normal result.

To make this explicit, 
we remark that the currents are 
conveniently expressed in terms of 
their 
electron numbers $n_s$ and $n_d$.
In effect, it is straightforward to show 
\(
 J_s^s =2 n_s e 
\)
from (\ref{JssequivJsphis}),
or more generally, we get it by a partial integration as follows.
\begin{eqnarray}
 J_s^s 
&=& 2e \sum_k 
{\bi u}\cdot {\bi v}_s(k)
\left(
-\frac{\partial f^0(\varepsilon_s(k)) }{\partial \varepsilon_s(k)}
\right)
{\bi u}\cdot {\bi k}
\label{Jss=-2e}
\\
&=& - 2e \sum_k 
{\bi u}\cdot \nabla_{\bi k}
\left(
{\bi u}\cdot {\bi k} f^0(\varepsilon_s(k))\right)
+ 2e \sum_k 
f^0(\varepsilon_s(k)).
\nonumber
\end{eqnarray}
The first term represents the contribution from 
the Brillouin zone boundary of the $k$ sum,
which vanishes when the states there are unfilled.
The second sum gives the result of the total number times $e$.
%
%
%
%
%
%
Similarly, we obtain 
\(
 J_d^d =2 n_d e 
\)
for the $d$ electron.
These results simply represent that 
the whole electrons are drifting all together,
as noted in the last section.
Hence, 
%
%
from (\ref{S=fracUss+Udd/T}) we get
\begin{equation}
S_0= \frac{
\displaystyle
{n_s} S_0^s+ {n_d} S_0^d
}{
\displaystyle{n_s} +{n_d}
}.
\label{S_0=diffuse}
\end{equation}
Especially, in the limit $n_d\gg n_s$, 
we obtain 
the enhanced diffusion thermopower 
$S_0\simeq S_0^d $ 
given in (\ref{S0^d}), which is wholly due to the $d$ electrons
carrying the spin fluctuations.

\subsection{Equilibrium effect}
\label{subsec:ee}

To the extent that we make use of an approximate expression \( J_d^d
\simeq 2 n_d e\) as above,
one may obtain $U_d^d \simeq 2 e_d$ correspondingly similarly,
where $e_d$ generally represents free energy of the $d$ electrons.
Then we obtain
\begin{equation}
S_0^d \simeq  {e_d}/(T n_d e).
\label{S0edTnde}
\end{equation}
This expression may be valuable as it 
is expressed in terms of the {\it equilibrium} quantities,
which have been vigorously investigated.
%
%
For example, one may have recourse to 
scaling argument for $e_d$.\cite{lrvw07}
We obtain $S_0^d  \propto T$ by $e_d \propto T^2$ normally, 
while at the 
QCP,
 $S_0^d \propto T\log T^{-1}$
according to $e_d \propto T^2\log T^{-1}$.
In terms of 
the electronic heat capacity $C$, 
one may substitute $e_d=CT $ to obtain
$S_0^d \simeq  C /(n_d e)$,
or 
\begin{equation}
 q\equiv \frac{e S}{ C }
= \frac{1}{n},
\label{qequiv es/c}
\end{equation}
where $n\equiv n_s+ n_d$ and
$S\simeq S_0 \simeq n_d S_0^d/ n $ under (\ref{s0dggs0s}).
For hole like carriers,
following as in (\ref{Jss=-2e}), 
we find that the number $n_d$ becomes negative
with the absolute value $|n_d|$ representing the hole number.
Thus our drag mechanism 
supports 
the material-independent universality in $q$
as revealed by Behnia $et$ $al$.\cite{BJF04}
This is contrasted with the explanation 
by resorting to dominant impurity scatterings.\cite{MK05}

To go further to investigate the next order contributions, 
we have to consider 
not only those originating from the equilibrium quantities,
which may be related to singular behaviour of the specific heat,
but also the {\it non-equilibrium} effect 
which manifest itself in linear response to an applied field.
The latter, though potentially important, has not been investigated before.
In the next section, we focus ourselves to 
such singular contributions which vanish at zero field ${\bi E}=0$.
We find similar temperature dependences
as that expected from the equilibrium effect through (\ref{S0edTnde}).



\section{\label{sec:std}Sub-leading corrections}

\subsection{Extra currents}


The effect of spin fluctuations on the single particle excitation
of conduction electron is described by a particle self-energy
$\Sigma({\bi k}, \varepsilon)$.
The dragged spin fluctuations 
bring about a similar effect
as those in equilibrium affect the thermodynamical properties.\cite{de66,bmf68}
We pay attention to the extra quasiparticle currents 
induced by the change of states at the Fermi level,
as they are expected to make dominant contributions.
%
%
We write an energy shift caused by a non-vanishing factor $\Phi^d_k$
as $\delta \varepsilon_s(k)$. Then the extra currents are given by 
\begin{equation}
{J}_s [ \Phi^d] =
2e \sum_k  {\bi u}\cdot {\bi v}_{s}(k)
 \frac{\partial f^0}{\partial
 \varepsilon_{s}(k)} 
\delta \varepsilon_{s}(k),
\label{Jsphid} 
\end{equation}
and 
\begin{equation}
{U}_s [ \Phi^d] =
2 \sum_k  {\bi u}\cdot {\bi v}_{s}(k)
(\varepsilon_{s}(k) -\mu)
 \frac{\partial f^0}{\partial \varepsilon_{s}(k)} 
\delta \varepsilon_{s}(k).
\label{Usphid} 
\end{equation}
The effective energy of the conduction electron
at the Fermi level
is given in terms of the real part of the self-energy ${\rm Re } \Sigma(k, \varepsilon)$
by 
\[
 \varepsilon^*_{s}(k) = 
\frac{\varepsilon_{s}(k) + 
{\rm Re } \Sigma (k,0)}{ 
\displaystyle 
1
 -\frac{\partial }{\partial \varepsilon}{\rm Re } \Sigma(k, 0)
 }.
\]
For the self-energy, 
we are interested in
those part
induced by the dragged spin fluctuations, 
which we denote as $\delta\left({\rm Re } \Sigma(k, 0)\right)$.
Thus
we have
\begin{equation}
\delta \varepsilon_{s}(k) \simeq 
\delta\left(
 {\rm Re } \Sigma (k,0) \right)
+(\varepsilon_{s}(k)-\mu) 
\frac{\partial }{\partial \varepsilon}
 \delta\left({\rm Re } \Sigma(k, 0)\right),
\label{deltavarepsksimeq}
\end{equation}
as we need 
$\delta \varepsilon_{s}(k)$ and $\delta\left({\rm Re } \Sigma(k, 0)\right)$
only to the linear order in $\Phi^d_k$.
The first and the second terms 
in (\ref{deltavarepsksimeq})
contribute mainly to
${J}_s [ \Phi^d]$ and ${U}_s [ \Phi^d]$, respectively.
In effect, we find
\begin{eqnarray}
\fl
{ U}_s [ \Phi^d] =
2 \sum_k  {\bi u}\cdot {\bi v}_{s}(k)
(\varepsilon_{s}(k) -\mu)^2
 \frac{\partial f^0}{\partial
 \varepsilon_{s}(k)} 
 \frac{\partial }{\partial \varepsilon}
\delta\left(
{\rm Re } \Sigma(k, 0)
\right)
\nonumber\\
\simeq 
2\langle
\left({\bi v}_{s}(k)\cdot {\bi u}\right)
 \frac{\partial }{\partial \varepsilon}
\delta\left(
{\rm Re } \Sigma(k, 0)
\right)
\rangle_{k_{F,s}}
\int_0^\infty
\rho_s(\varepsilon) {\rm d}\varepsilon
(\varepsilon_{s,k} -\mu)^2
 \frac{\partial f^0}{\partial
 \varepsilon_{s,k}} 
\label{simeqleftvskuright}\\
=
- \frac{2\pi^2}{3}\rho_{F,s}  I'(0) T^2,
\label{UsPhid}
\end{eqnarray}
where $I'(0)$ is the derivative at $\varepsilon=0$ of 
\begin{equation}
I(\varepsilon)=
\langle
{\bi u}\cdot{\bi v}_{s}(k)
\delta \left(
{\rm Re } \Sigma({ k}, \varepsilon)
\right)
\rangle_{k_{F,s}}.
\label{Ivarepsilondef}
\end{equation}
The angular bracket 
in (\ref{Ivarepsilondef})
represents the average over the Fermi surface.
In (\ref{simeqleftvskuright}), $\rho_s(\varepsilon)$ 
is the 
DOS
per spin of the $s$  electron,
and $\rho_{F,s}=\rho_s(\varepsilon_{F,s})$.
Furthermore, we used 
\[
\int_0^\infty
\rho_s(\varepsilon) {\rm d}\varepsilon
(\varepsilon_{s,k} -\mu)^2
 \frac{\partial f^0}{\partial
 \varepsilon_{s,k}} 
=-
 \frac{\pi^2}{3}\rho_{F,s}  T^2.
\]
Similarly as (\ref{UsPhid}), we obtain
\begin{equation}
{J}_s [ \Phi^d]= -2 \rho_{F,s}I(0), 
\label{JsPhid}
\end{equation}
using $I(\varepsilon)$ in (\ref{Ivarepsilondef}).
As we find $I(0)$ is insignificant, 
a correction to 
the thermopower due to the $s$ electrons affected by 
the spin fluctuations is given by 
\begin{equation}
\Delta S_s 
=\frac{{U}_s[\Phi^d]}{T({ J}_s^s+{ J}_d^d)}
=
- \frac{\pi^2}{3 e }
\frac{
\rho_{F,s}  I'(0) 
}{ n_s+n_d}T.
\label{DeltaS}
\end{equation}

\subsection{Self-energy}


We employ the self-energy in which
a spin fluctuation excitation is emitted at one vertex and absorbed at the other one. 
It is given by
\begin{equation}
 \Sigma({\bi k}, \varepsilon_n)
=-\frac{3}{2}J^2 
T \sum_{\varepsilon_n'} \sum_{k'}
{G}({\bi k}',\varepsilon_n')
\chi
(
{\bi k}-{\bi k}',
\varepsilon_n-\varepsilon_n')
\end{equation}
where $\varepsilon_n=(2n+1)\pi T$ and 
$\varepsilon_n'=(2n'+1)\pi T$ are the 
fermion Matsubara frequencies,
${G}({\bi k},\varepsilon_n)$ is 
the temperature Green's function for the $s$ electron,
and $
\chi
({\bi q},\omega_n)$ 
is related to the 
$d$ electron susceptibility 
$
\chi
({\bi q},\omega)$
at the imaginary frequency $\omega = {\rm i}\omega_n$, where $\omega_n=2n\pi T$
 is the boson Matsubara frequency.
By an analytic continuation, we obtain
the following relation for 
the retarded functions,
denoted below with the subscript $R$, which are 
analytic in the upper half plane
of the complex frequencies;
\begin{eqnarray}
\fl
{\rm Re} \Sigma_R({\bi k}, \varepsilon)= 
-
\frac{3}{2}J^2 
\sum_{k'} 
\int^\infty_{-\infty}
\frac{{\rm d}\omega}{2\pi} 
 {\rm Im }
G_R({\bi k}',\omega) {\rm Re} \chi_R(
{\bi k}-{\bi k}',
\varepsilon-\omega)
\tanh \frac{\omega}{2T}
\nonumber\\
-
\frac{3}{2}J^2 
 \sum_{k'} 
 \int^\infty_{-\infty}
\frac{{\rm d}\omega}{2\pi} 
{\rm Re} G_R({\bi k}',\varepsilon-\omega) {\rm Im } 
\chi_R
(
{\bi k}-{\bi k}',
\omega)
\coth \frac{\omega}{2T}.
\label{SigmaR}
\end{eqnarray}
To obtain the effect of $\Phi^d_k$,
we substitute
\(
\chi_R(q,\omega) =\chi_{\rm drag}(q,\omega +{\rm i}\delta)
\) from (\ref{chiphiq=}).
Hence the shift $\delta \left(
{\rm Re} \Sigma_R({\bi k}, \varepsilon)
\right)$
is obtained from 
 (\ref{SigmaR})
by substituting 
$
\frac{\partial 
\chi_R(q,\omega ) 
}{\partial \omega}
\Phi_q^d 
$ in place of 
$
\chi_R(q,\omega) $.
For $G_R({\bi k}',\omega)$, we use a free propagator 
\(
G_R({\bi k},\omega) ={1}/({\omega - \xi_k +  {\rm i}\delta}),
\)
where \( \xi_k =\varepsilon_s(k)-\mu\).
Owing to 
 \( {\rm Im} G_R({\bi k}',\omega) 
 = - \pi \delta (\omega -  \xi_{k'})\) and (\ref{chi+-qomega})
for 
$
\chi_R({\bi q},\omega)
$,
the first term of (\ref{SigmaR}) gives
\begin{eqnarray}
\fl {
\sum_{k'} 
\int^\infty_{-\infty}
\frac{{\rm d}\omega}{2\pi} 
 {\rm Im }
G_R({\bi k}',\omega) 
\frac{\partial}{\partial \varepsilon}
{\rm Re}\chi_R({\bi k}-{\bi k}',\varepsilon-\omega)
\tanh \frac{\omega}{2T}
\Phi_{k-k'}^d }
\nonumber\\
=
-\frac{1}{2} 
\sum_{q} 
\frac{\partial}{\partial \varepsilon}
{\rm Re}
\chi_R({\bi q},\varepsilon-\xi_{k-q})
\left(1- 2f^0(\xi_{k-q})\right)\Phi_q^d.
\nonumber
\end{eqnarray}
As this give only a convergent result, we neglect this part.
Using (\ref{phiks=ukphiqd=uq}) for $\Phi_q^d$,
for (\ref{Ivarepsilondef}) we find
\begin{equation}
\fl I(\varepsilon)=
-
\frac{J^2}
{2(2\pi)^4}\int {\rm d}{\bi q} \int^\infty_{-\infty}
\frac{{\rm d}\omega }{\varepsilon-\omega 
+ {\bi v}_{s}(k)\cdot {\bi q}
}
\frac{\partial
{\rm Im } \chi_R({\bi q},\omega)
}{\partial \omega}\coth \frac{\omega}{2T}
\langle{\bi v}_s(k_{F,s})\cdot{\bi q}\rangle_{k_{F,s}},
\end{equation}
where we substituted
$\xi_{k-q} \simeq  - {\bi v}_{s}(k_{F,s})\cdot {\bi q}$,
which holds in the important integral region of small $|{\bi q}|$.
Integrating over the angle between ${\bi v}_{s}(k_{F,s})$ and ${\bi q}$,
we obtain 
\begin{eqnarray}
\fl
I(\varepsilon) 
=
-
\frac{J^2}{2(2\pi)^3}
\int_0^{2k_{F,s}} q^2{\rm d}q
\int^\infty_{-\infty}{\rm d}\omega 
\left(
2-
\frac{\varepsilon-\omega}{{v}_{F,s}q}
\log
\left|
\frac{
\varepsilon-\omega+{v}_{F,s}q 
}
{
\varepsilon-\omega-{v}_{F,s}q }
\right|
\right)
\nonumber\\
\times \frac{\partial
 {\rm Im } \chi_R({q},\omega)
}{\partial \omega}\coth \frac{\omega}{2T}.
\label{defIeps}
\end{eqnarray}
In the parenthesis,
only those terms odd in $\omega$ 
contribute to the integral over $\omega$.
Hence we find
\(I(0) =0,\) 
and the leading term in
$|\omega/(v_{F,s} q)|$ gives
\begin{eqnarray}
\fl
I'(0)\simeq
-
\frac{J^2}{2\pi^3{v}_{F,s}^2}
\int_0^{2k_{F,s}} {\rm d}q
\int^\infty_0
\omega {{\rm d}\omega }
\frac{\partial
 {\rm Im } \chi_R({q},\omega)
}{\partial \omega}\coth \frac{\omega}{2T}
\\
= 
-
\frac{J^2 k_{F,d}\varepsilon_{F,d}}{{2\pi^3}{{v}_{F,s}^2}}
\int_0^{2k_{F,s}/k_{F,d}}
 {\rm d}\bar{q}
 \int^\infty_0
\bar{\omega} {{\rm d}\bar{\omega} }
\frac{\partial
 {\rm Im } \chi_R(\bar{q},\bar{\omega})
}{\partial \bar{\omega}}\coth \frac{
\varepsilon_{F,d}\bar{\omega}
}{2T}.
\end{eqnarray}
To put in this expression, we may write the susceptibility 
in (\ref{chi+-qomega})
as
\[
\frac{\partial}{\partial \bar{\omega}}
{\rm Im } \chi_R({q},{\omega}) =
\frac{\partial}{\partial \bar{\omega}}
\left(
\frac{\bar{\omega}}{\bar{q}}
\frac{A}{\left(
\bar{\kappa}^2+
\bar{q}^2
\right)^2
+
 \left(
C
\bar{\omega}/\bar{q}
\right)^2
}
\right)
\]
for 
\(
 \bar{\omega} <  2\bar{q},
\)
where 
$ A={36 \pi \rho_{F,d}}/{\bar{U}^2}$,
$C =3 {\pi}$, and
\begin{equation}
 \bar{\kappa}^2= 
 12 K_0^2/{\bar{U}}
= 12 (1-{\bar{U}})/{\bar{U}}.
\end{equation}
We find
\begin{eqnarray}
\fl
 I'(0)
=
-
\frac{J^2 k_{F,d}\varepsilon_{F,d} A}{{\pi^3}{{v}_{F,s}^2}}
\left(
{\cal I}_0 +{\cal I}(T)
\right)
=
-
36
\left(
\frac{\bar{J}}{\bar{U}}
\right)^2
\left(
\frac{k_{F,d}}{k_{F,s}}
\right)^4
\left(
{\cal I}_0 +{\cal I}(T)
\right),
\label{I'0=calI0IT}
\end{eqnarray}
where 
\(
 \bar{J}\equiv \rho_{F,s} J,
\)
\begin{equation}
\fl
{\cal I}_0=
 \frac{1}{2}
\int_0^{2k_{F,s}/k_{F,d}}
\frac{{\rm d}\bar{q}}{{\bar{q}}}
 \int^{2\bar{q}}_0
\bar{\omega} {{\rm d}\bar{\omega} }
\frac{\partial}{\partial \bar{\omega}}
\left(
\frac{
{\bar{\omega}}
}{\left(
\bar{\kappa}^2+
\bar{q}^2
\right)^2
+
 \left(
C
\bar{\omega}/\bar{q}
\right)^2
}
\right),
\end{equation}
and
\begin{equation}
\fl
{\cal I}(T)=\int_0^{2k_{F,s}/k_{F,d}}
\frac{{\rm d}\bar{q}}{{\bar{q}}}
 \int^{2\bar{q}}_0
\bar{\omega} {{\rm d}\bar{\omega} }
\frac{\partial}{\partial \bar{\omega}}
\left(
\frac{
{\bar{\omega}}
}{\left(
\bar{\kappa}^2+
\bar{q}^2
\right)^2
+
 \left(
C
\bar{\omega}/\bar{q}
\right)^2
}
\right)
n^0(\varepsilon_{F,d}\bar{\omega}).
\label{calI}
\end{equation}
The former ${\cal I}_0$ is the part independent of temperature $T$, 
while 
the temperature dependence in the latter ${\cal I}(T)$ arises
from the Bose factor $n^0(\omega)$.
In particular, for $\bar{\kappa}=0$, we obtain
\begin{eqnarray}
{\cal I}_0 
&=&
 \frac{1}{4}
\int_0^{(2k_{F,s}/k_{F,d})^2}
{{\rm d}\bar{q}^2}
 \int^{2\bar{q}}_0
\bar{\omega} {{\rm d}\bar{\omega} }
\frac{\partial}{\partial \bar{\omega}}
\left(
\frac{
{\bar{\omega}}
}{
\bar{q}^6
+
 \left(
C\bar{\omega}
\right)^2
}
\right)
\nonumber\\&=&
\int_0^{(2k_{F,s}/k_{F,d})^2}{{\rm d}\bar{q}^2}
\left(
\frac{1 }{\bar{q}^4+(2C)^2 }
+
 \frac{1}{8C^2}
 \log \frac{\bar{q}^4}{\bar{q}^4 + (2C)^2
}
\right)
\nonumber\\
&=&
 \frac{(k_{F,s}/k_{F,d})}{4C^2}
 \log \frac{( k_{F,s}/k_{F,d})^2}{
(k_{F,s}/k_{F,d})^2+ C^2}.
\end{eqnarray}

\subsection{\label{sec:calIT}Temperature dependence: ${\cal I}(T)$}

To obtain an explicit expression for 
 the temperature dependent part ${\cal I}(T)$,
we adopt
an approximation to set
\begin{equation}
 n^0(\omega)=
\left\{
\begin{array}{ll}
\displaystyle {T}/{\omega}, \quad& \omega< cT\\
 0, & \omega>cT\\
\end{array}
\right.
\label{n0omega=om<cT}
\end{equation}
where $c$ is a constant of order unity
(cf. (\ref{constc0.928})).
Consequently, we obtain
\begin{equation}
{\cal I}(T)= 
{\cal I}_a(T)+{\cal I}_b(T),
\end{equation}
where 
\begin{equation}
\fl
{\cal I}_a(T) = 
\frac{1}{2}
\left(\frac{T}{\varepsilon_{F,d}}\right)^2
\int_0^{\bar{q}_0^2}
{{\rm d}\bar{q}^2}
 \int^{2\bar{q} \varepsilon_{F,d}/T}_0
{{\rm d}u }
\frac{\partial}{\partial u}
\left(
\frac{{u}}{
\bar{q}^2\left(
\bar{\kappa}^2+
\bar{q}^2
\right)^2
+
 \left(
 C Tu
/\varepsilon_{F,d}
\right)^2
}
\right),
\label{IaTdef}
\end{equation}
and 
\begin{equation}
\fl
{\cal I}_b(T) = 
\frac{1}{2}
\left(\frac{T}{\varepsilon_{F,d}}\right)^2
\int_{\bar{q}_0^2}^{(2k_{F,s}/k_{F,d})^2}
{{\rm d}\bar{q}^2}
 \int^{c
}_0
{{\rm d}u }
\frac{\partial}{\partial u}
\left(
\frac{{u}}{
\bar{q}^2\left(
\bar{\kappa}^2+
\bar{q}^2
\right)^2
+
 \left(
 C Tu
/\varepsilon_{F,d}
\right)^2
}
\right).
\label{IbTdef}
\end{equation}
Here we introduced
a characteristic scale
for the normalized momentum, 
\begin{equation}
 \bar{q}_0  \equiv \frac{c T}{2\varepsilon_{F,d}}.
\label{q0def} 
\end{equation}
We may take the limit $\bar{\kappa}=0$ 
for (\ref{IaTdef}) to obtain
\begin{eqnarray}
{\cal I}_a(T) 
\simeq 
\frac{2T}{\varepsilon_{F,d}}
\int_0^{\bar{q}_0}
\frac{{{{\rm d}\bar{q}}
}}{
\bar{q}^4
+
 \left(
2C
\right)^2
}
\simeq 
\frac{ {\bar{q}_0}T}{
 2C^2\varepsilon_{F,d}}
=
c
\left(\frac{T}{2C\varepsilon_{F,d}}\right)^2.
\end{eqnarray}
On the other hand, for (\ref{IbTdef}), 
we obtain
\begin{equation}
{\cal I}_b(T) = 
\frac{c}{2}
\left(\frac{T}{\varepsilon_{F,d}}\right)^2
\int_{\bar{q}_0^2}^{(2k_{F,s}/k_{F,d})^2}
\frac{
{{\rm d}\bar{q}^2}
}{
\bar{q}^2\left(
\bar{\kappa}^2+
\bar{q}^2
\right)^2
+
 \left(
2C \bar{q}_0
\right)^2
},
\label{Ib(T)}
\end{equation}
for which the main contribution
comes from around the lower limit of the integral.
Let us discuss two cases depending on the relative size of 
 $\bar{\kappa}$ and $\bar{q}_0$, separately.



First we consider the case $\bar{\kappa}/\bar{q}_0 \gg 1$,
which is the low temperature limit 
for $\bar{\kappa}>0$.
In this case, we obtain
\begin{eqnarray}
{\cal I}_b(T)
&\simeq &
\frac{c}{2}
\left(\frac{T}{\varepsilon_{F,d}}\right)^2
\frac{1}{\bar{\kappa}^4}
\int_{\bar{q}_0^2}^{(2k_{F,s}/k_{F,d})^2}
\frac{
{{\rm d}\bar{q}^2}
}{
\bar{q}^2
+
 \left(
2C \bar{q}_0
/\bar{\kappa}^2
\right)^2
}
\nonumber\\
&\simeq&
\frac{c}{2}
\left(\frac{T}{\varepsilon_{F,d}}\right)^2
\frac{1}{\bar{\kappa}^4}
\log \frac{
{(k_{F,s}/k_{F,d})^2}
+
 \left(
C \bar{q}_0
/\bar{\kappa}^2
\right)^2
}{
 \left(
C \bar{q}_0
/\bar{\kappa}^2
\right)^2
}.
\end{eqnarray}
In terms of a characteristic temperature of spin fluctuations defined by
\begin{equation}
\bar{T} \equiv 
\frac{\varepsilon_{F,d} \bar{\kappa}^2}{C}
=\frac{4 \varepsilon_{F,d} K_0^2}{\pi{\bar{U}}},
\label{Tsfdef}
\end{equation}
we find
\begin{equation}
{\cal I}_b(T)
\simeq
\frac{c}{2C^2}
\left(\frac{T}{
 \bar{T}}\right)^2
\log \frac{
{(2 k_{F,s}/k_{F,d})^2}
+
 \left(c T/\bar{T}\right)^2
}{
 \left(c T/\bar{T}\right)^2
}.
\label{IbT=c2C2log}
\end{equation}
In the literature, 
a spin fluctuation temperature,
\begin{equation}
{T}_{\rm sf}
=\varepsilon_{F,d} K_0^2
={\varepsilon_{F,d}}({1-\bar{U}}),
\end{equation}
is commonly used as well.
Indeed we have $\bar{T} \simeq {T}_{\rm sf} $
for $\bar{U}\simeq 1$.
%
Lastly, in the quantum critical limit $\bar{\kappa}/\bar{q}_0 \ll 1$,
 we obtain
\begin{eqnarray}
{\cal I}_b(T) 
%
&\simeq&
\frac{1}{2}
\left(\frac{T}{\varepsilon_{F,d}}\right)^2
\int_{\bar{q}_0^2}^{(2k_{F,s}/k_{F,d})^2}
\frac{
{{\rm d}\bar{q}^2}
}{
\bar{q}^6
+
 \left(
2C \bar{q}_0
\right)^2
}
\nonumber\\&\simeq&
 \frac{\sqrt{3}{\pi}}{9 (2C\bar{q}_0)^{4/3}}
\left(\frac{T}{\varepsilon_{F,d}}\right)^2
=
 \frac{{\pi}}{3\sqrt{3} (C {c })^{4/3}}
\left(\frac{T}{\varepsilon_{F,d}}\right)^{2/3}.
\label{Ib(T)c}
\end{eqnarray}

\subsection{\label{sec:summary}Results}

We may neglect ${\cal I}_a(T)$ against ${\cal I}_b(T)$,
for $\bar{T}\ll \varepsilon_{F,d}$.
For (\ref{DeltaS}),
we obtain 
\begin{equation}
\Delta S_s
\simeq 
\Delta S_0^s + \Delta S_s(T),
\end{equation}
where
\begin{equation}
\Delta S_0^s 
=
 \frac{ \rho_{F,s}  T
}{3 e( n_s+n_d) }
\left(
\frac{\bar{J}}{\bar{U}}
\right)^2
\left(
\frac{k_{F,d}}{k_{F,s}}
\right)^3
 \log \frac{( k_{F,s}/k_{F,d})^2}{
(k_{F,s}/k_{F,d})^2+ (3\pi)^2},
\end{equation}
and
\begin{equation}
 \Delta S_s(T)
=
 \frac{12\pi^2}{ e( n_s+n_d) }
\left(
\frac{\bar{J}}{\bar{U}}
\right)^2
\left(
\frac{k_{F,d}}{k_{F,s}}
\right)^4
\rho_{F,s}{\cal I}_b(T)
  T.
\label{DeltaSs=24pi2}
\end{equation}
The former $\Delta S_0^s$ to modify the $T$ linear term may be effectively neglected,
while the latter $ \Delta S_s(T)$ gives a sub-leading correction.
At the low temperature $T\ll \bar{T}$,
with (\ref{IbT=c2C2log}), we get
\begin{eqnarray}
\fl \Delta S_s(T)
&\simeq &
 \frac{2 }{3 e( n_s+n_d) }
\left(
\frac{\bar{J}}{\bar{U}}
\right)^2
\left(
\frac{k_{F,d}}{k_{F,s}}
\right)^4
\rho_{F,s}
  T
\left(\frac{T}{
 \bar{T}}\right)^2
\log \frac{
{(2 k_{F,s}/k_{F,d})^2}
+
 \left( T/\bar{T}\right)^2
}{
 \left( T/\bar{T}\right)^2
},
\label{Delta S_s(T)}
\end{eqnarray}
where we set $c\simeq 1$ for simplicity (instead of (\ref{constc0.928})).
In the opposite limit, from
(\ref{Ib(T)c}), we obtain
\begin{eqnarray}
 \Delta S_s(T)
&\simeq&
 \frac{4\pi^{5/3}}{{3}^{11/6} e( n_s+n_d) }
\left(
\frac{\bar{J}}{\bar{U}}
\right)^2
\left(
\frac{k_{F,d}}{k_{F,s}}
\right)^4
\rho_{F,s}
  T
 \frac{}{ }
\left(\frac{T}{\varepsilon_{F,d}}\right)^{2/3}.
\label{Deltasst.QC}
\end{eqnarray}
%



In the same manner as $U_s[\Phi^d]$ discussed above, 
one can think of an additional heat current
$\Delta U_d[\Phi^d]$
caused by the intraband many-body effect due to 
the on-site repulsion $U$ in the $d$ band.
Formally, the corresponding results are obtained
straightforwardly by replacing  
$k_{F,s}$, $\rho_{F,s}$, and $3J^2/2$ in the above results
by $k_{F,d}$, $\rho_{F,d}$, and $U^2$,
respectively, i.e., 
\begin{equation}
 \Delta S_d(T)
\simeq 
 \frac{4 }{9 e( n_s+n_d) }
\rho_{F,d}
  T
\left(\frac{T}{
 \bar{T}}\right)^2
\log \frac{
4
+
 \left( T/\bar{T}\right)^2
}{
 \left( T/\bar{T}\right)^2
},
\label{DeltaSdT}
\end{equation}
in place of (\ref{Delta S_s(T)}).
The results are modified in some ways in generalizing the model.
The constant of 4 in the logarithm of (\ref{DeltaSdT})
stems from $(2k_{F,d}/k_{F,d})^2$, 
where $2k_{F,d}$ sets the upper cutoff 
for the momentum $q$ of spin fluctuations.
If we should have set a cutoff parameter $q_c$ differently,
the factor should be replaced by $\bar{q}_c^2$, where
$\bar{q}_c \equiv q_c/k_{F,d}$.
Moreover, if we had assumed a phenomenological coupling $g$ between electrons and 
spin fluctuations instead of $U$, 
the results will be reduced by a factor of $(g/U)^2$.

\section{\label{sec:disscs}Discussion: comparison with experiment}

\begin{figure}
\begin{center}
\includegraphics{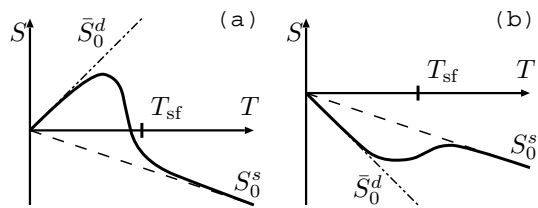}
\caption{\label{fig1} 
The effect of spin-fluctuation drag on the thermopower $S(T)$ is
 schematically shown.
The bold lines are drawn to interpolate the two linear relations, 
$S(T)=\bar{S}_0^d$ and $\bar{S}_0^s$.
The low temperature $\bar{S}_0^d$ in (\ref{Sdrag simeq nd}) 
is due to drag of those heavy electrons pertaining to the spin
 fluctuations,
while
the normal diffusion term 
$S\simeq {S}_0^s$ due to light conduction electrons appear
at high temperature $T\gg \bar{T}\simeq T_{\rm sf}$, a spin-fluctuation temperature.
The left (a) is for $\bar{S}_0^d>0$, while (b) for $\bar{S}_0^d<0$.
The latter
is compared with
${\rm RCo_2}$ (R=Sc, Y and Lu) by
Gratz $et$ $al$.\cite{grbbmg95,gratz97,gm01} 
}
\end{center}\end{figure}

To compare the theoretical result $S(T)$ with experiment, 
some assumptions like the free dispersions in
(\ref{varepsk}) and (\ref{varepdk})
should not be taken literally. 
In particular, the $T$-linear terms $S_0^i$ ($i=s,d$) for $S_0$
would be able to have either positive or negative sign,
depending on the energy dependence of the 
respective DOS
at the Fermi level,
while
the relation $|S_0^d|\gg |S_0^s|$ will always hold true for their relative magnitudes.
%
Therefore, as the leading effect at low temperature, 
we generally expect
an enhanced $T$-linear term,
\begin{equation}
S_0
\simeq 
\bar{S}_0^d\equiv  \frac{
\displaystyle
 {n_d}
}{
\displaystyle{n_s} +{n_d}
} S_0^d,
\label{Sdrag simeq nd}
\end{equation}
unless $n_s\gg n_d$.
Effectively, 
this term is indistinguishable
from the diffusion term contribution,
as discussed below (\ref{S0^d}).
It is indeed due to the drag current of the heavy $d$ electrons.
Without drag,
we recover the conventional result $S\simeq S_0\simeq S_0^s$
of the diffusion thermopower due to the conduction electrons.
We expect that the latter holds true 
at high temperature $T\agt \bar{T}$
where the $s$-$d$ scatterings
become too weak to sustain the $d$ electron drag. 
Therefore, 
it is reasonably expected that
we should find some structure in the temperature dependence of the thermopower
$S(T)$ around $T\alt \bar{T}$,
which is brought about by the crossover between the $T$-linear terms 
with different magnitudes of coefficients.
This is schematically shown in figure~\ref{fig1}.

Takabatake $et$ $al$.\cite{tmnhmsus06}
have shown experimentally by applying the magnetic field of 15T that
an S-shaped structure in $S(T)$ of ${\rm CaFe_4Sb_{12}}$
observed at low $T< \bar{T}\simeq 50$K is suppressed 
to yield a normal $T$-linear diffusion term.
This is consistent with our result for
$\bar{S}_0^d<0$, $S_0^s>0$
and $|\bar{S}_0^d/S_0^s|\simeq 8$.
In this case, the conduction band for $S_0^s$ consists mainly of
 5$p$ states of antimony.
Moreover, they have shown that
the temperature dependence of 
the spin-fluctuation contribution $\Delta S=S-S_0^s$
is not monotonic.
To explain this theoretically
goes beyond the scope of this paper, as it requires us to
solve the transport equations concretely.
Similarly known before were the low temperature minima in
the thermopower of ${\rm RCo_2}$ (R=Sc, Y and Lu),
which had been stressed by Gratz $et$ $al$.\cite{grbbmg95,gratz97,gm01} 
as the experimental evidence of paramagnon drag.
Their results can be compared with our result 
for $\bar{S}_0^d \ll  S_0^s< 0$ in figure~\ref{fig1} (b).

%
%

On the correction terms, 
it is generally expected that
the $d$ electron contribution
$\Delta S_d$ will become more important than $\Delta S_s$
 when the $d$ electron current becomes relevant indeed.
As discussed in section~\ref{subsec:ee},
we have to consider two sources of contributions,
one due to the equilibrium effect 
and the other due to the non-equilibrium effect in section~\ref{sec:std}.
Interestingly, we find that both give the same temperature dependence, 
\(
S(T)= \alpha T + \beta  T^3
\log T^{-1}
\)
away from the QCP. 
Nevertheless, we notice an important difference.
While we observe $\beta\propto 1/K_0^4$
from the results of the last section,
$\beta$ expected from a correction term in (\ref{S0edTnde}) has 
an extra factor of $1/K_0^2$.\cite{de66}
This means that the equilibrium effect becomes more important.
We suspect that this would hold true at the QCP too,
although there has been no definite calculation deriving
the corresponding free energy correction $\propto T^{5/3}$ 
in accordance with our result.

In any case, 
we remark that the relative magnitude of 
the electron numbers $n_s$ and $n_d$ may have an effect on the
correction terms,
the sign of which will depend on the factor $
en= e(n_s+ n_d)$, 
that is, the direction of the net current.
In most cases where the model applies, 
the current carrier in the heavy-electron band
will be  hole like.
%
Moreover,
we generally expect that $|n_s|$ will not exceed $|n_d|$, 
or 
the net current would be hole-like, $e(n_s+ n_d) >0$. 
Accordingly, $\Delta S_i >0$ (cf. (\ref{qequiv es/c})).
This is consistent with a model calculation of 
the spin fluctuation effect on the resistivity, where
Jullien $et$ $al.$\cite{jbc74,cij78}
pointed out the important role of the parameter $\xi = k_{F,c}/k_{F,d}$ 
on the transport properties of spin fluctuations systems.
We observe the dependence in 
our results of (\ref{Delta S_s(T)}) and (\ref{Deltasst.QC}).
To compare their numerical results with experiments,
they should choose $\xi\le 1$ generically, that is, $|n_s|\alt |n_d|$.

\begin{figure}
\begin{center}
\includegraphics{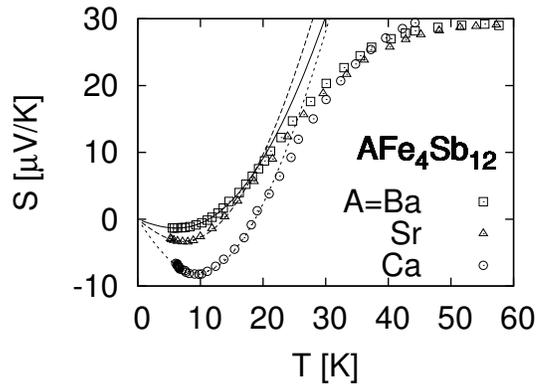}
\caption{\label{fig} 
The points are the experimental data of $S(T)$ for $\rm
 AFe_4Sb_{12}$ $\rm (A=Ba, Sr, Ca)$.\cite{mhittm05}
The lines are the least squares fits
by the theoretical expression in (\ref{STLSF})
with the parameters 
given in table \ref{tab:table}.
}
\end{center}
\end{figure}

To conclude, 
let us fit the low-temperature experimental data for $S(T)$ of
$\rm AFe_4Sb_{12}$ $\rm (A=Ba, Sr, Ca)$ 
reported by Matsuoka $et$ $al.$\cite{mhittm05}
with
\begin{equation}
S(T)= \alpha T + \beta  T
\left(\frac{T}{ \bar{T}}\right)^2
\log \frac{
\delta 
+
 \left( T/\bar{T}\right)^2
}{
 \left( T/\bar{T}\right)^2
},
\label{STLSF} 
\end{equation}
where $\alpha$, $\beta$,  $\bar{T}$ and $\delta$ are regarded as
parameters.
In table~\ref{tab:table}, we present the fitting parameters 
obtained for $\delta =4$ 
by the least squares fits of the low temperature part of the data for
$T\alt 17$K $(< \bar{T})$.
The results are shown in figure~\ref{fig}, 
along with the experimental data points.
We find that $\alpha$'s do not depend much on the other parameters, 
and the ratios of $\beta$ and $\bar{T}$ between
materials are nearly independent of $\delta$.
The relatively large values for $\alpha$ and $\beta$ will be more properly 
ascribed to the heavier $d$ band 
than to the conduction band, in accordance with our result.
Note that these coefficients are
susceptible to the equilibrium effect of mass enhancement,\cite{de66,bmf68} 
which we did not take into account explicitly (cf. section~\ref{subsec:ee}).\footnote{
Owing to prevalent anharmonic phonons in this skutterudite system,
it may not be a simple matter to extract the electronic contribution 
from the observed specific heat coefficient $\gamma$,
which does not depend sensitively on the divalent ion A.\cite{mhittm05}
}
The positive $\beta $ implies that
the net current is in the hole-like direction.
The relative material dependence of $\beta$ in table \ref{tab:table} 
may be qualitatively compared with the observed static uniform susceptibility
$\chi_0 \propto \rho_d/K_0^2$,
that is, 
$\chi_0({\rm BaFe_4Sb_{12}}): 
\chi_0({\rm SrFe_4Sb_{12}}): 
\chi_0({\rm CaFe_4Sb_{12}})\simeq 1:1.6:2.5
$.

\begin{table}
\caption{\label{tab:table}
Parameters to fit the thermopower 
of $\rm AFe_4 Sr_{12}$ 
in figure~\ref{fig}.
}
\begin{indented}
\lineup
\item[]
\begin{tabular}{@{}lccc}
\br
& $\alpha$ {[$\rm \mu V/K^2$]} & $\beta$ {[$\rm \mu V/K^2$]} & $\bar{T}$ {[K]}\\
\mr
$\rm BaFe_4 Sb_{12}$ & $-0.35$ & 1.5 & 48\\ 
$\rm SrFe_4 Sb_{12}$ & $-0.74$ & 1.6 & 37\\ 
$\rm CaFe_4 Sb_{12}$ & $-1.4\0$ & 3.7 & 58\\ 
\br\end{tabular}
\end{indented}
\end{table}

\ack
The author is very grateful to Eiichi Matsuoka 
for the original data of \cite{mhittm05}.

\appendix

\section{\label{sec:formaltransporttheory}Formal transport theory}

The formal expressions 
for resistivity and thermopower
referred to 
in the main text are 
derived by adapting a general variational method of Ziman,\cite{ziman60}
according to which 
$\Phi^s$ in (\ref{f=f0-Phik})
and $\Phi^d$ in (\ref{chiphiq=}) 
are regarded as variational trial functions.
Below we substitute $\eta_i \Phi^i$ for $\Phi^i$ ($i=s,d$), and 
take variation with respect to the arbitrary parameters $\eta_i$.

On the one hand, 
the microscopic entropy production rate
corresponding to (\ref{fkscattLinearD})
is given by
\begin{eqnarray}
 \dot{S}_{\rm scatt}&=& \frac{1}{T^2}
\sum_{k,q}
 \left(
\eta_s
 \Phi_k^s 
- 
\eta_s\Phi_{k+q}^s
+ \eta_d \Phi_q^d 
\right)^2
{\cal P}^{k+q}_{kq} 
\nonumber
\\
&&\equiv \frac{1}{T} 
\sum_{i,j=s,d}
P_{ij} \eta_i \eta_j.
\label{Sscatt=}
\end{eqnarray}
The components of the matrix $P_{ij}$ defined in (\ref{Sscatt=}) 
are explicitly given by 
\begin{eqnarray}
P_{ss}  &=&\frac{1}{T}\sum_{k,q}
{\cal P}^{k+q}_{kq}  \left(
 \Phi_k^s 
- 
\Phi_{k+q}^s
\right)^2,
\label{Pss}
\\
P_{sd}=P_{ds}  &=&\frac{1}{T}\sum_{k,q}
{\cal P}^{k+q}_{kq}  \left(
 \Phi_k^s 
- 
\Phi_{k+q}^s
\right) \Phi_q^d ,
\label{Psd}
\\ 
P_{dd}  &=&\frac{1}{T}\sum_{k,q}
{\cal P}^{k+q}_{kq}  
 (\Phi_q^d) ^2.
\label{Pdd}
\end{eqnarray}
In (\ref{Sscatt=}),
not only emission of a paramagnon 
corresponding to (\ref{fkscattLinearD}),
but the reverse absorption process 
is also taken into account.
In the special case
of the full drag without Umklapp processes,
there holds the relation $\Phi_{k+q}^s - \Phi_k^s =\Phi_q^d$
by (\ref{phiks=ukphiqd=uq}), so that
we get the following identities,
\begin{equation}
 P_{ss}=P_{dd} =-P_{sd}.
\label{Pss=Pdd=-Psd}
\end{equation}



On the other hand, the macroscopic entropy production is 
given by
\begin{equation}
 \dot{S}_{\rm macro}= 
 \frac{{\bi J} \cdot {\bi E}}{T} +
{\bi U}
\cdot  \nabla \frac{1}{T}.
\end{equation}
In the linear response regime, the electric current ${\bi J}$ and 
the heat current ${\bi U}$ are written as
\begin{eqnarray}
 {\bi J} &=&
\eta_s {\bi J}[\Phi^s]
+\eta_d {\bi J}[\Phi^d],
\label{J=etasd}
\\
 {\bi U} &=&
\eta_s {\bi U}[\Phi^s]
+\eta_d {\bi U}[\Phi^d],
\label{U=etasd}
\end{eqnarray}
where 
${\bi J}[\Phi^i]$ denotes the current flow
caused by $\Phi^i$ ($i=s,d$), i.e., 
${\bi J}[\Phi^i]$ formally represents the part of the total current 
which depends linearly on  $\Phi^i$.
${\bi U}[\Phi^i]$ is similarly defined.
In general, these currents
have different functional forms.
It is remarked that ${\bi J}[\Phi^i]$ is not to be identified with
the current in the $i$ band.
Owing to the interband interaction,
the distribution shift $\Phi^i$ in the $i$ band can induce a current in
the other band.

The variational parameters $\eta_i$ are determined 
so as to maximize $\dot{S}_{\rm scatt}$ 
after equating $\dot{S}_{\rm scatt}$ and $\dot{S}_{\rm macro}$.\cite{ziman60}
Substituting the solutions into 
(\ref{J=etasd}) and (\ref{U=etasd}),
we obtain
\begin{equation}
 {\bi J} 
=
\sum_{i,j=s,d} {\bi J}[\Phi^i]
(P^{-1})_{ij}
\left(
{{\bi J}[\Phi^j]
 \cdot {\bi E}} 
-\frac{1}{T} {\bi U}[\Phi^j]
\cdot  \nabla T
\right),
\label{J=sumijsd=J}
\end{equation}
and
\begin{equation}
 {\bi U} 
=\sum_{i,j=s,d}
{\bi U}[\Phi^i]
(P^{-1})_{ij}
\left(
{
{\bi J}[\Phi^j]
 \cdot {\bi E}} 
-\frac{1}{T} 
{\bi U}[\Phi^j]
\cdot  \nabla T
\right),
\label{U=sumijsd=U}
\end{equation}
where $(P^{-1})_{ij}$ is the inverse matrix of $P_{ij}$.
For definiteness, 
let the applied field ${\bi E}$ and $\nabla T$
be in the direction of a unit vector ${\bi u}$.
In an isotropic system,
or in cubic symmetry,
the results are expressed with the magnitudes
$J[\Phi^i] ={\bi J}[\Phi^i] \cdot {\bi u}$
and
$U[\Phi^i] ={\bi U}[\Phi^i] \cdot {\bi u}$.
From (\ref{J=sumijsd=J}), 
we obtain the electrical conductivity, 
\[
 \sigma =
\sum_{l,m=s,d} { J}[\Phi^l]
(P^{-1})_{lm}
{ J}[\Phi^m].
\]
The resistivity $R=\sigma^{-1}$ is given by 
\begin{equation}
 R=
R_0(T)\frac{
\displaystyle
{1 - \frac{P_{sd}P_{ds}}{P_{ss} P_{dd}}}
}{
\displaystyle
1+
\left(
\frac{{ J}[\Phi^d]}{{ J}[\Phi^s]}
\right)^2
\frac{P_{ss}}{P_{dd}}
},
\label{R=R01-frapsd}
\end{equation}
where 
\begin{equation}
 R_0 (T)= \frac{P_{ss}}{({J}[\Phi^s])^2}.
\label{R0T=fracpssA}
\end{equation}
The latter, given in (\ref{R0T=fracpss}),
is the resistivity that we obtain when we have no spin-fluctuation drag.
In fact, this is the central formula to explain
an enhanced resistivity 
of a spin fluctuation system 
due to 
normal scattering processes 
with long-lived spin fluctuations.\cite{ml66,rice68,mathon68}
%
%
According to (\ref{R=R01-frapsd}),
the $d$ electron drag modifies
the resistivity in two ways.
First, we note that the numerator in (\ref{R=R01-frapsd})
vanishes in the full drag case, (\ref{Pss=Pdd=-Psd}).
This represents physically that
a finite resistivity is brought about 
only with those scattering processes
which can degrade the total net current.
%
%
%
On the basis of a more realistic model,
a proper treatment of Umklapp scattering processes 
could make the numerator a non-vanishing factor of order unity.
Secondly, the positive factor in the denominator 
has the effect of suppressing the resistivity.
This is due to an additional drag current of the $d$ electrons.
When fully dragged, 
the $d$ electrons carry $n_d/n_s$ times as large current as 
the $s$ electrons,
where $n_d/n_s$ is the ratio of the electron densities.
In general, this would not be negligible quantitatively, 
and it might be so even qualitatively.

From the condition of no heat flow ${\bi U}=0$ for (\ref{U=sumijsd=U}),
we obtain the Seebeck coefficient,
\begin{equation}
S =
\frac{1}{T}\frac{
\displaystyle
\sum_{l,m=s,d} { J}[\Phi^l]
(P^{-1})_{lm}
{ U}[\Phi^m]
}{
 \displaystyle
\sum_{l,m=s,d} { J}[\Phi^l]
(P^{-1})_{lm}
{ J}[\Phi^m]
}.
\end{equation}
From this we can obtain the result for the the full drag case of (\ref{Pss=Pdd=-Psd})
formally as a special limit. 
It is expressed simply
by the ratio of the total energy current to the
total momentum current as
\begin{equation}
 S =
\frac{1}{T}
\frac{
{ U}[\Phi^s]
+{U}[\Phi^d]
}{
{ J}[\Phi^s]
+{ J}[\Phi^d]
}.
\label{S=frac1TUs+Ud/Js+Jd}
\end{equation}
Indeed, 
the simple result
in this limit is straightforwardly generalized to many-band models.
It is owing to this simple property that
we investigated this limit devotedly in the main text.

On the other side,
the case without drag is obtained for
${U}[\Phi^d]={ J}[\Phi^d]=0$ as
\begin{equation}
 S_0^s =
\frac{1}{T}
\frac{
{ U}[\Phi^s]
}{
{ J}[\Phi^s]
},
\label{S0s=1/Tfrac}
\end{equation}
as presented in (\ref{S0sT}).
As a matter of fact,
the no-drag results of (\ref{R0T=fracpssA}) and
(\ref{S0s=1/Tfrac}) are directly derived
without taking $\Phi^d$ into account from the beginning.

\section{\label{sec:IbT}
Temperature dependence of
${\cal I}(T)$
 at $\bar{\kappa}=0$}

To evaluate ${\cal I}(T)$ in (\ref{calI}), 
we made the approximation as given in (\ref{n0omega=om<cT}).
We obtained (\ref{Ib(T)c}) for ${\cal I}_b(T)$ in
(\ref{IbTdef}),
which signifies the main correction term of
$\Delta S\propto T^{5/3}$
in the quantum critical regime. 
The exponent 5/3 is the same as 
for the resistivity.\cite{mathon68}
%
The derivation in section~\ref{sec:calIT}
indicates that 
the important contributions come from 
$\omega \simeq T$.
In effect, this is the upper limit 
 of the $\omega$ integral
for $\bar{q}\agt \bar{q}_0$,
and the high-energy cutoff is naturally provided by 
the Bose factor $n^0(\omega)$ in the integrand,
without employing the approximation in (\ref{n0omega=om<cT}).
With this in mind, 
we can obtain the result for $\bar{\kappa}=0$
directly by transforming the integral 
and taking the limits for 
the bounds of integration
as follows.
\begin{eqnarray}
\fl
{\cal I}_b(T) 
=
\frac{1}{2 \varepsilon_{F,d}^2}
\int_{\bar{q}_0^2}^{2k_{F,s}/k_{F,d}}
{{\rm d}\bar{q}^2}
 \int^{2\bar{q}}_0
{\omega} {{\rm d}{\omega} }
\frac{\partial}{\partial{\omega}}
\left(
\frac{
{{\omega}}
}{
\bar{q}^6
+
 \left(
C
{\omega}/\varepsilon_{F,d}
\right)^2
}
\right)
n^0(\omega)
\nonumber\\
\simeq 
\frac{1}{2 \varepsilon_{F,d}^2}
\int_{\bar{q}_0^2}^{2k_{F,s}/k_{F,d}}
{{\rm d}\bar{q}^2}
 \int^{\infty}_0
{\omega} {{\rm d}{\omega} }
\frac{\partial}{\partial{\omega}}
\left(
\frac{
{{\omega}}
}{
\bar{q}^6
+
 \left(
C
{\omega}/\varepsilon_{F,d}
\right)^2
}
\right)
n^0(\omega)
\nonumber\\
=
-\frac{T^{2/3}}{2 \varepsilon_{F,d}^2}
\int_{\bar{q}_0^2/T^{2/3}}^{2k_{F,s}/k_{F,d}/T^{2/3}}
{{\rm d}v 
}
 \int^{\infty}_0
\frac{
{u} {{\rm d}u }
}{
v^3
+
 \left(
C u
/\varepsilon_{F,d}
\right)^2
}
\frac{\partial}{\partial u}
\left(\frac{u}{{\rm e}^u-1}
\right)
\nonumber\\
\simeq  
-\frac{T^{2/3}}{2 \varepsilon_{F,d}^2}
\int_0^\infty
{{\rm d}
v 
}
 \int^{\infty}_0
\frac{
{u}{{\rm d}u }
}{
v^3
+
 \left(C u/\varepsilon_{F,d}\right)^2
}
\frac{\partial}{\partial u}
\left(\frac{u}{{\rm e}^u-1}
\right)
\nonumber\\
=  
-
\frac{\pi}{3\sqrt{3} C^{4/3}}
\left(\frac{T}{\varepsilon_{F,d}}\right)^{2/3}
 \int^{\infty}_0
 {{\rm d}u }
u^{-1/3}
\frac{\partial}{\partial u}
\left(\frac{u}{{\rm e}^u-1}\right)
\nonumber\\
= 
1.10
\frac{\pi}{3\sqrt{3}C^{4/3}}
\left(\frac{T}{\varepsilon_{F,d}}\right)^{2/3}.
\label{Ib(T)ex}
\end{eqnarray}
By comparing (\ref{Ib(T)ex}) and (\ref{Ib(T)c}), 
we obtain
\(
{c }^{-4/3} \simeq 1.10
\),
or 
\begin{equation}
{c} 
\simeq 0.928.
\label{constc0.928}
\end{equation}

\section*{References}

\bibliography{main}

\end{document}